\documentclass[11pt]{iopart}
\usepackage{epsfig}
\usepackage{subfigure}
\usepackage{iopams}
\usepackage{amssymb}
\usepackage{graphics}
\usepackage{graphicx}
\usepackage{psfrag}
\usepackage{showidx}
\usepackage{color}

\begin{document}

\newcommand{\BEQ}{\begin{equation}}     
\newcommand{\BEA}{\begin{eqnarray}}
\newcommand{\BD}{\begin{displaymath}}
\newcommand{\EEQ}{\end{equation}}       
\newcommand{\EEA}{\end{eqnarray}}
\newcommand{\ED}{\end{displaymath}}
\newcommand{\nn}{\nonumber}
\newcommand{\E}{{\rm e}}
\newcommand{\eps}{\varepsilon}          
\newcommand{\vph}{\varphi}              
\newcommand{\vth}{\vartheta}            
\newcommand{\D}{{\rm d}}                
\newcommand{\II}{{\rm i}}               
\newcommand{\arcosh}{{\rm arcosh\,}}    
\newcommand{\erf}{{\rm erf\,}}          
\newcommand{\erfc}{{\rm erfc\,}}        
\newcommand{\wit}[1]{\widetilde{#1}}    
\newcommand{\wht}[1]{\widehat{#1}}      
\newcommand{\lap}[1]{\overline{#1}}     
\newcommand{\demi}{\frac{1}{2}}         
\newcommand{\rar}{\rightarrow}          
\newcommand{\gop}{\wht{\phi}_{\vec{0}}} 

\renewcommand{\vec}[1]{{\bf{#1}}}       
\newcommand{\matz}[4] 
     {\mbox{${\begin{array}{cc} #1 & #2 \\ #3 & #4 \end{array}}$}}
\newcommand{\iint}{\int\!\!\int}
\newcommand{\textc}{\textcolor{black}}
\newcommand{\textr}{\textcolor{black}}
\newcommand{\<}{\langle}
\renewcommand{\>}{\rangle}

\def\g{\gamma}
\def\r{\rho}
\def\w{\omega}
\def\wo{\w_0}
\def\wp{\w_+}
\def\wm{\w_-}
\def\t{\tau}
\def\av#1{\langle#1\rangle}
\def\pf{P_{\rm F}}
\def\pr{P_{\rm R}}
\def\F#1{{\cal F}\left[#1\right]}
\renewcommand{\sc}{\MakeUppercase}

\title[Boundary crossover in non-equilibrium growth processes]{Boundary crossover 
in semi-infinite non-equilibrium growth processes}

\author{Nicolas Allegra, Jean-Yves Fortin and Malte Henkel}
\address{Groupe de Physique Statistique, 
D\'epartement de Physique de la Mati\`ere et des Mat\'eriaux, 
Institut Jean Lamour (CNRS UMR 7198), Universit\'e de Lorraine Nancy, 
B.P. 70239, F -- 54506 Vand{\oe}uvre-l\`es-Nancy Cedex, France}
\ead{nicolas.allegra@univ-lorraine.fr, jean-yves.fortin@univ-lorraine.fr, malte.henkel@univ-lorraine.fr} 

\date{\today}

\begin{abstract}

\textc{The growth of stochastic interfaces in the vicinity of a boundary and the 
non-trivial crossover towards the behaviour deep in the bulk is analysed. 
The causal interactions of the interface with the boundary lead to a roughness 
larger near to the boundary than deep in the bulk.}  
This is exemplified in the semi-infinite Edwards-Wilkinson model in one dimension, 
both from its exact solution and numerical simulations, as well as from simulations on the semi-infinite 
one-dimensional Kardar-Parisi-Zhang model. The 
non-stationary scaling of interface heights and widths is analyzed and 
a universal scaling form for the local height profile is proposed. 
 
\end{abstract}

   
\maketitle

Improving the understanding of growing interfaces continues \cite{Bara95} 
as a widely fascinating topic of
statistical physics, with a large variety of novel features 
still being discovered. When considering the growth
of interfaces, the situation most commonly analysed is the one of 
a spatially infinite substrate. Alternatively, if
substrates of a finite size are studied, periodic boundary conditions are used. 
Then, boundary effects need not be taken
into account and the scaling of the interface fluctuation can be described in terms of the 
standard Family-Vicsek ({\sc{fv}}) \cite{Fami85} 
description. 

\textr{Starting from the height $h(t,x)$, the usual definition of the 
interface roughness, across the system, is \cite{Bara95}}
\BEQ\label{def1}
w_{{\rm exp},L}^2(t):={\left\< {\,\overline{\left[h(t,x)-\overline{h(t,x)}\right]^2}} \,\right\>}=
{\left\< \overline{h^2(t,x)}\right\>}-\left\<\overline{h(t,x)}^{\:2}\right\>,
\EEQ
\textr{where $\overline{X}=\overline{X}(t)=L^{-1}\int_L\!\D x\, X(t,x)$ is the spatial average in 
a system of total spatial size $L$. 
The ensemble average (over the random noise and/or over all possible 
realisations) is denoted by $\<X(t,x)\>$.
A very commonly studied situation is a finite system of linear 
size $L$ with {\em periodic} boundary conditions. Then, the {\sc{fv}} scaling form 
is \cite{Fami85}}
\BEQ\label{eq2}
\!w_{{\rm exp},L}(t) = t^{\beta} f\left( L t^{-1/z}\right) \: , \:
f(u) \sim \left\{ \begin{array}{ll} u^{\alpha}       & \mbox{\rm  ; if $u\ll 1$} 
\\ 
                                    \mbox{\rm const} & \mbox{\rm  ; if $u\gg 1$}
\end{array}\right.
\EEQ
\textr{
with the growth exponent $\beta=\alpha/z$. Recently, clear experimental examples 
with exponents of the Kardar-Parisi-Zhang universaliy class (see below) have 
been found in turbulent liquid crystals \cite{Take11} and growing cancer cells 
\cite{Huergo12}. If one furthermore uses a spatially translation-invariant initial 
condition (e.g. a flat initial surface), then both the average surface profile 
$\langle h(t,x)\rangle$ and its local width, which we define as
\footnote{\textr{In principle, the Fourier transform of a correlator 
such as $w^2(t,x)$ can be measured in scattering experiments at finite momentum 
$q$. In the limit $q\rightarrow 0$ we would recover $w_{{\rm 
exp},L}^2(t)$.}}
}
\BEQ \label{eq1}
w^2(t,x) := \left\langle \left[h(t,x) - \left\langle 
h(t,x)\right\rangle\right]^2\right\rangle 
= \left\langle h^2(t,x)\right\rangle - \left\langle h(t,x)\right\rangle^2,
\EEQ
\textr{ 
are space-translation-invariant, hence independent of the location $x$. 
Then, the width $w^2(t,x)\to 
w_{L}^2(t):=\overline{w^2(t,x)}=\overline{\< 
h(t,x)^2\>}-\overline{\<h(t,x)\>^2}$ can be formally rewritten as a spatial average. 
At first sight, the two definitions of the bulk
roughness, $w_{{\rm exp},L}(t)$ and $w_{L}(t)$, should be different. However, since for a
system with spatial translation-invariance one has 
$\overline{\<h(t,x)\>^2}=\left\<\overline{h(t,x)}\right\>^2$, they are identical, 
viz. $w_{{\rm exp},L}^2(t)=w_{L}^2(t)$, and the explicit average over space in the definition of 
$w_{L}^2(t)$ merely serves to reduce stochastic noise. 
}

\begin{figure}
\centerline{\epsfxsize=3.0in\ \epsfbox{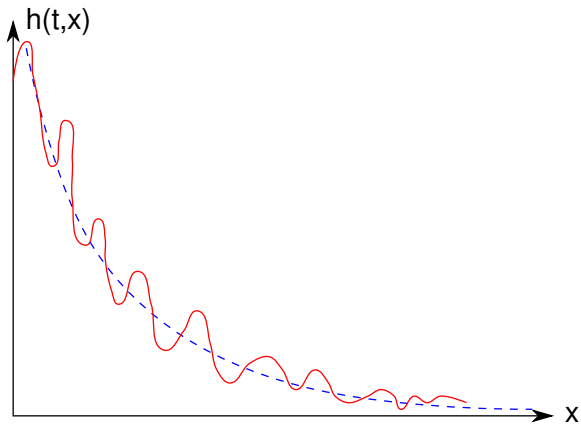}}
\caption[fig0]{Schematic noise-averaged interface $\langle h(t,x)\rangle$ (dashed line) 
and the actual fluctuations around it (full line), in the presence of a boundary at 
$x=0$. \label{fig0}}
\end{figure}

Here, we are interested in how a boundary in the substrate may 
affect the properties of the interface. \textr{For a system on the half-line $x\geq 0$ 
with a boundary at $x=0$, space-translation-invariance is broken and both $\langle h(t,x)\rangle$
and $w(t,x)$ depend on the distance $x$ from the boundary, see figure~\ref{fig0}. 
Still, one expects that deep in the bulk $x\gg 1$, the width 
$w(t,x)$ should converge towards the bulk roughness 
$w^2(t,x)\stackrel{x\to\infty}{\rar}w_{\infty}^2(t)=w_{{\rm exp},\infty}^2(t)$.
For finite values of $x$, however, the precise properties of the width $w(t,x)$ 
will depend on the precise boundary conditions not contained in global quantities such as 
$w_{L}(t)$ and $w_{{\rm exp},L}(t)$.}
\textr{We shall use (\ref{eq1}) 
to measure the fluctuations
of the interface around its position-dependent, ensemble-average value 
$\left\langle h(t,x)\right\rangle$.} 
This question has since long ago been studied in the past, see \cite{Wolf90,Krug91}. 
These studies usually begin by
prescribing some fixed boundary conditions and 
then proceed to analyse the position-dependent interface, often
through the height profile. 
In this work, we start from the situation where particles are deposited on a bounded substrate 
and first ask how the deposition rules become modified in the 
vicinity of the boundary as compared to deep in the bulk. 
Generically, bulk models of particle-deposition select first a site on which one 
attempt to deposit a particle, often
followed by a slight redistribution of the particle in the vicinity of the initially selected site,
where the details of these rules lead to the numerous 
recognised universality classes \cite{Bara95,Halp95,Krug97}. 
If one conceives of the boundary as a hard wall which the particles cannot penetrate,  
those particles which would have to leave the system are kept on the boundary 
site. Restricting this study for simplicity to 
a semi-infinite system in $d=1$ space dimensions, this suggests the {\em 
boundary condition $H_0(t)\ge H_1(t)$}, where $H_i(t)$ is the local height on 
the site $i\geq 0$ such that the boundary occurs at $i=0$.

In contrast to earlier studies \cite{Wolf90,Krug91}, 
it turns out that a continuum description of the interface requires 
a careful re-formulation of the known bulk models such that the height 
$h(t,x)$ near to the boundary must be determined self-consistently
(throughout, we shall work in the frame moving with the mean interface velocity). 
\textc{Technically, this can be achieved by performing a Kramers-Moyal 
expansion on the master 
equation which describes the lattice model defined above. 
The calculations are rather lengthy and will be reported in detail elsewhere. In the continuum limit,} 
this leads to a {\em non-stationary height profile} \textc{$\langle 
h(t,x)\rangle = 
t^{1/\gamma} \Phi( x^z t^{-1})$}, 
with the new scaling relation
$\gamma=z/(z-1)$, although stationarity is kept deep in the bulk, where $x\to\infty$ and $\Phi(\infty)=0$. 
Unexpected behaviour is also found for the width profile $w(t,x)$, see below.

Another motivation of this work comes from the empirical observation that 
{\sc{fv}}-scaling must be generalised in that 
{\em global} and {\em local} fluctuations with different values of 
$\beta$ are to be distinguished \cite{Rama06}. Furthermore,
the experimentally measured values of $\beta$ 
\cite{{Nasci11},{Yim09},{Duerr03},{Mata08},{Gong11},{Kim10},{Cord09}} 
are larger than those expected in many simple model systems \cite{Bara95,Halp95,Krug97}. 
In certain cases, these enhanced values of
$\beta$ are experimentally observed together with a grainy, 
{\em faceted morphology of the interfaces} \cite{Nasci11,Yim09,Duerr03,Cord09} and
furthermore, a cross-over in the effective value of 
$\beta$ from small values at short times to larger values at longer times is seen \cite{Nasci11}. 
While this  might suggest that some new exponent should be introduced, this is contradicted by 
renormalization-group ({\sc{rg}}) arguments in bulk systems without disorder or 
long-range interactions \cite{Lopez05}.

We study the influence of a substrate boundary on a growing 
interface by analysing the simplest case of a single boundary in a
semi-infinite system, with the condition $H_0(t)\ge H_1(t)$. 
It is left for future work to elucidate any possible direct relevance for anomalous roughening. 
First, we shall study the semi-infinite $1d$ Edwards-Wilkinson ({\sc{ew}}) 
model 
\cite{Edwa82}, whose bulk behaviour is well-understood from an exact solution.
We shall write down a physically correct Langevin equation in semi-infinite space which includes this new
kind of boundary contribution. Its explicit exact solution, 
of both the height profile as well as the site-dependent
interface width, will be seen to be in agreement with a large-scale Monte Carlo 
({\sc{mc}}) simulations. It turns out that there exists
a surprisingly large intermediate range of times where the model crosses over to a new fixed point,
with an effective and non-trivial surface growth exponent $\beta_{1,{\rm eff}}>\beta$, 
which is qualitatively analogous to the experiments cited above. 
However, at truly large times, \textc{typically above the diffusion time}, the 
system converts back to the
{\sc{fv}} scaling, as expected from the bulk {\sc{rg}} \cite{Lopez05}.
Second, in order to show that these observations 
do not come from the fact that the Langevin
equation of the {\sc{ew}} model is linear, 
we present numerical data for a model in the universality class of the
semi-infinite $1d$ Kardar-Parisi-Zhang ({\sc{kpz}}) equation \cite{Kard86}. 
We find the same qualitative results as for the
{\sc{ew}} class, with modified exponents. 
\textc{We point out that} the time scales needed to see these cross-overs are considerably larger 
than those studied in existing experiments.

\renewcommand{\thesubfigure}{(\roman{subfigure})}
\begin{figure}[]
\begin{center}
\subfigure[]
{
\includegraphics[height=5cm,keepaspectratio,clip]{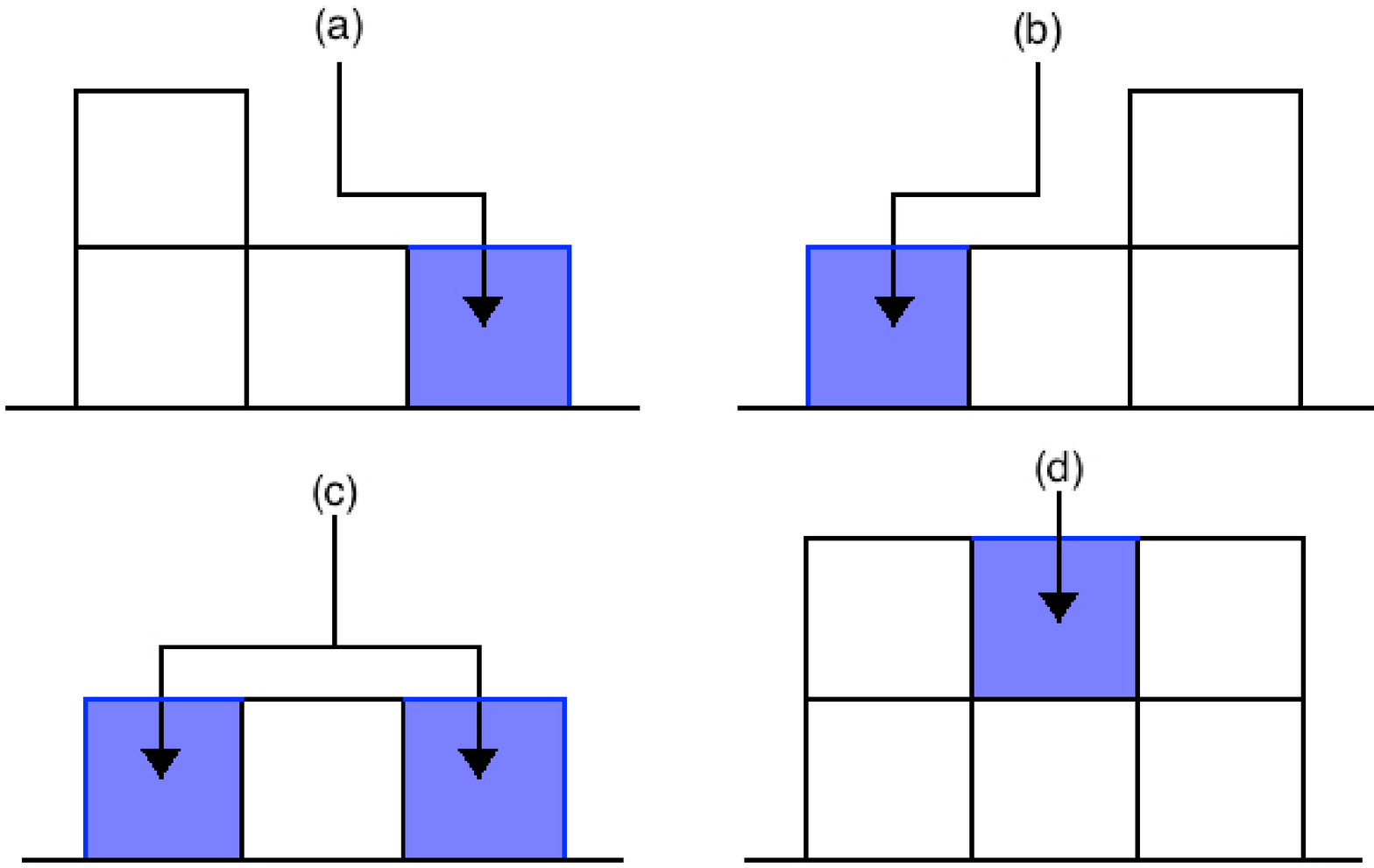}
\label{fig_bulk}
}
\subfigure[ ]
{
\includegraphics[height=4cm,keepaspectratio,clip]{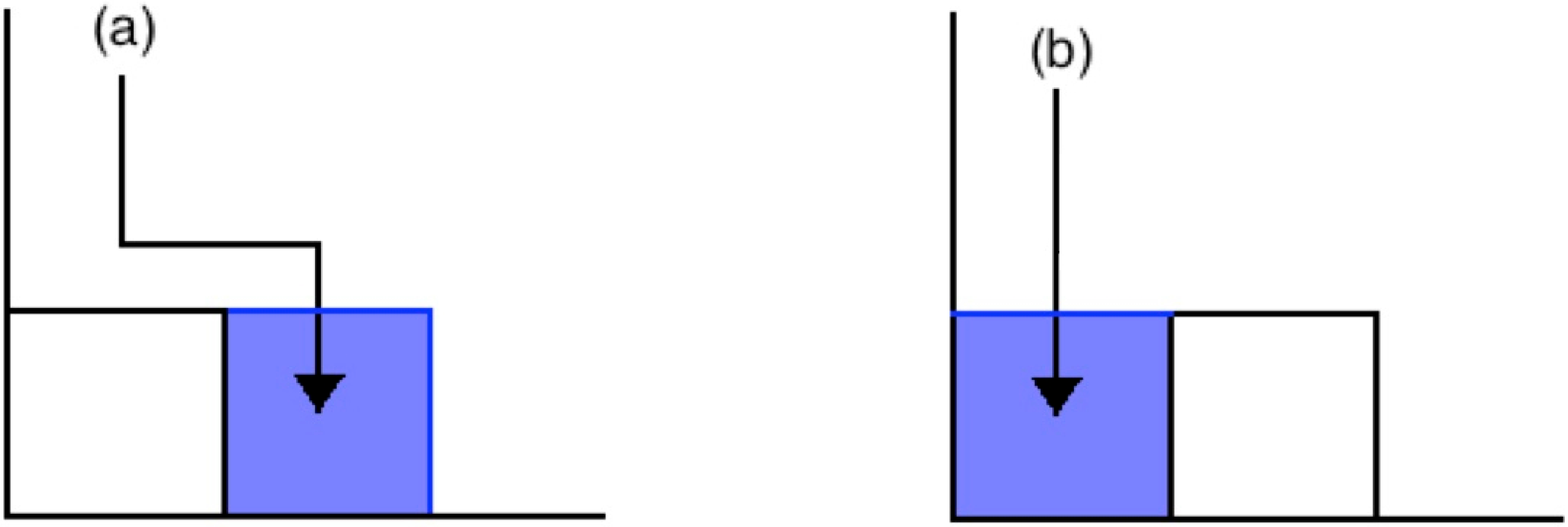}
\label{fig_bord}
}
\end{center}
\caption[fig_proc]{\textc{(i) Family algorithm~\cite{Fami86} in the bulk, the 
particle 
stays where it falls if the height of the two neighbouring sites is higher (d), 
or it diffuses to the right or left with the same probability if the 
neighbours have lower or equal height (c). Finally, the particle diffuses 
to the right (left) if the neighbour on the left (right) has higher height, see 
(a) and (b) respectively. 
(ii) Modification of the Family algorithm on the boundary, the particle 
diffuses to the right (a) if the height of the neighbour on the right is 
lower or equal, and stays where it falls otherwise (b).
\label{fig_proc} }}
\end{figure}
\renewcommand{\thesubfigure}{(\alph{subfigure})}

\hspace{-0.5truecm}\underline{\it Edwards-Wilkinson model.} In the following we study a microscopic process 
which should reproduce the $1d$ semi-infinite {\sc{ew}} equation, \textc{see figure~\ref{fig_proc}(i)}. 
A particle incident
on a site $i$ of a linear substrate of size $L$ remains there only if its height $H_{i}$ is 
less or equal to the heights of its two nearest neighbours. If only one
nearest-neighbour site has a lower height, deposition is 
onto that site, but if both nearest-neighbour heights are lower than that of the 
original site, the deposition site is chosen randomly between the two
lower sites with the same probability. In {\sc{mc}} simulations, we have taken 
$L = 10^{4}$ and all the 
data have been averaged over $2\cdot 10^{4}$ samples. This microscopic process 
\cite{Fami86} associated with periodic boundary condition is known to 
belong to the {\sc{ew}} universality class.
We modified the process at the boundary, \textc{see figure~\ref{fig_proc}(ii)}, 
by imposing that {\em the height of the 
first site must be higher than the one on the second site}, in order to simulate 
an infinite potential at the origin. This apparently rather weak boundary condition will be seen
to lead to significantly new and robust behaviour near to the boundary.

A continuum Langevin equation can be found starting from a master equation for 
the microscopic process  \cite{Vve93,Chu05}. However these methods 
may suffer from several problems, like the incomplete determination of the 
parameters and divergence in the continuum limit when the elementary space step 
is taken to zero. 
\textc{A Kramers-Moyal expansion on the master equation of the discrete model 
illustrated in figure~\ref{fig_proc} leads} 
in the continuum limit to the Langevin equation, 
defined in the 
half-space $x\geq 0$
\BEQ \label{eq3}
\left(\partial_t -\nu \partial_x^2\right)h(t,x) -\eta(t,x) = \nu\left(\kappa_1 + 
\kappa_2 h_{1}(t)\right)\delta(x) .
\EEQ
This holds in a co-moving frame with the average interface deep in the bulk \textc{(such that $\langle 
h(t,\infty)\rangle=0$)}. Furthermore, $\nu$ is the diffusion constant,
 $\kappa_1$ can be considered as an external 
{\em source} \textc{at the origin and one defines 
$h_{1}(t):=\left.\partial_x h(t,x)\right|_{x=0}$. Indeed, taking the noise 
average of \eref{eq3} and performing a space integration over the semi-infinite 
chain, we find that $\nu^{-1}\partial_t \int_0^{\infty}h(t,x)\D 
x=\kappa_1+(\kappa_2-1)h_{1}(t)$. We take $\kappa_2=1$ since this corresponds 
to the simplest condition of a constant boundary current 
$\nu\kappa_1 \delta(x)$.} The centered Gaussian 
noise has the variance $\left\langle \eta(t,x)\eta(t',x')\right\rangle=2\nu 
T\delta(t-t')\delta(x-x')$, where $T$ is an effective temperature. Parameters 
$\nu$, $\kappa_1$, and $T$ are material-dependent constants and 
initially, the interface is flat $h(0,x)=0$. With respect to the well-known 
description of the bulk behaviour \cite{Edwa82}, the
new properties come from the boundary terms on the r.h.s. of eq.~(\ref{eq3}). 
In contrast to earlier studies \cite{Wolf90}, the local slope 
$h_1(t)$ is not 
{\it a priori} given, but must be found self-consistently.
In what follows, we choose units such that $\nu=1$. 
A spatial Laplace transformation leads to
\BEA \label{eq4}
&&h(t,x) \!= \!\frac{1}{4\sqrt{\pi}} \!\int_{{x^2}/{4t}}^{\infty} 
\!\!\frac{\D v\,\E^{-v}}{v^{3/2}} 
\left[ x \kappa_1 +2v h_0\left(t-\frac{x^2}{4v}\right)\right] + \zeta(t,x) ,
\\
&&\zeta(t,x)=\int_0^t\frac{\D\tau}{\sqrt{4\pi(t-\tau)}}\int_0^{\infty}
\D x'\eta(\tau ,
x')\E^{-(x-x')^2/4(t-\tau)},
\nn
\EEA

where $h_0(t):=h(t,0)$. 
\textc{The modified noise $\zeta$ is related to noise $\eta$ in 
terms of its Laplace transform $\overline{\zeta}(t,p) := \int_0^{\infty}\!\D 
x\: \E^{-p x}\zeta(t,x) = \int_0^t \!\D \tau\: \E^{p^2(t-\tau)} 
\overline{\eta}(\tau,p)$.}
It remains to determine the function $h_0(t)$ self-consistently. Expanding 
eq.~(\ref{eq4}) to the first non-trivial order in $x$, we obtain by 
identification
$h_0(t) = 2\left[\kappa_{1}\pi^{-1/2}\sqrt{t}+\zeta(t,0)\right]$. Introducing the scaling 
variable $\lambda :=x^2/4t$, the 
height profile and 
height fluctuation can be cast into the scaling form
\BEA\label{eq7}
&&\!\!\!\!\langle h(t,x) \rangle= \sqrt{t}\, \Phi(\lambda) =
\sqrt{t}\,\frac{2\kappa_1}{\pi}\left[ \E^{-\lambda} 
-\sqrt{\pi\lambda\,}\,\erfc \sqrt{\lambda}\right],
\\ 
&&\!\!\!\!h(t,x)-\langle h(t,x) \rangle =\int_{\lambda}^{\infty} 
\!\!\frac{\E^{-v}\D v}{\sqrt{\pi v}}\: \zeta\left(t-\frac{\lambda 
t}{v},0\right) 
+\zeta\left(t,x\right).
\nn
\EEA
\begin{figure}[tb]
\begin{center}
\subfigure[ ]
{
\includegraphics[height=5cm,keepaspectratio,clip]{allegra1_fig1.eps}
\label{fig1}
}
\subfigure[ ]
{
\includegraphics[height=5.4cm,keepaspectratio,clip]{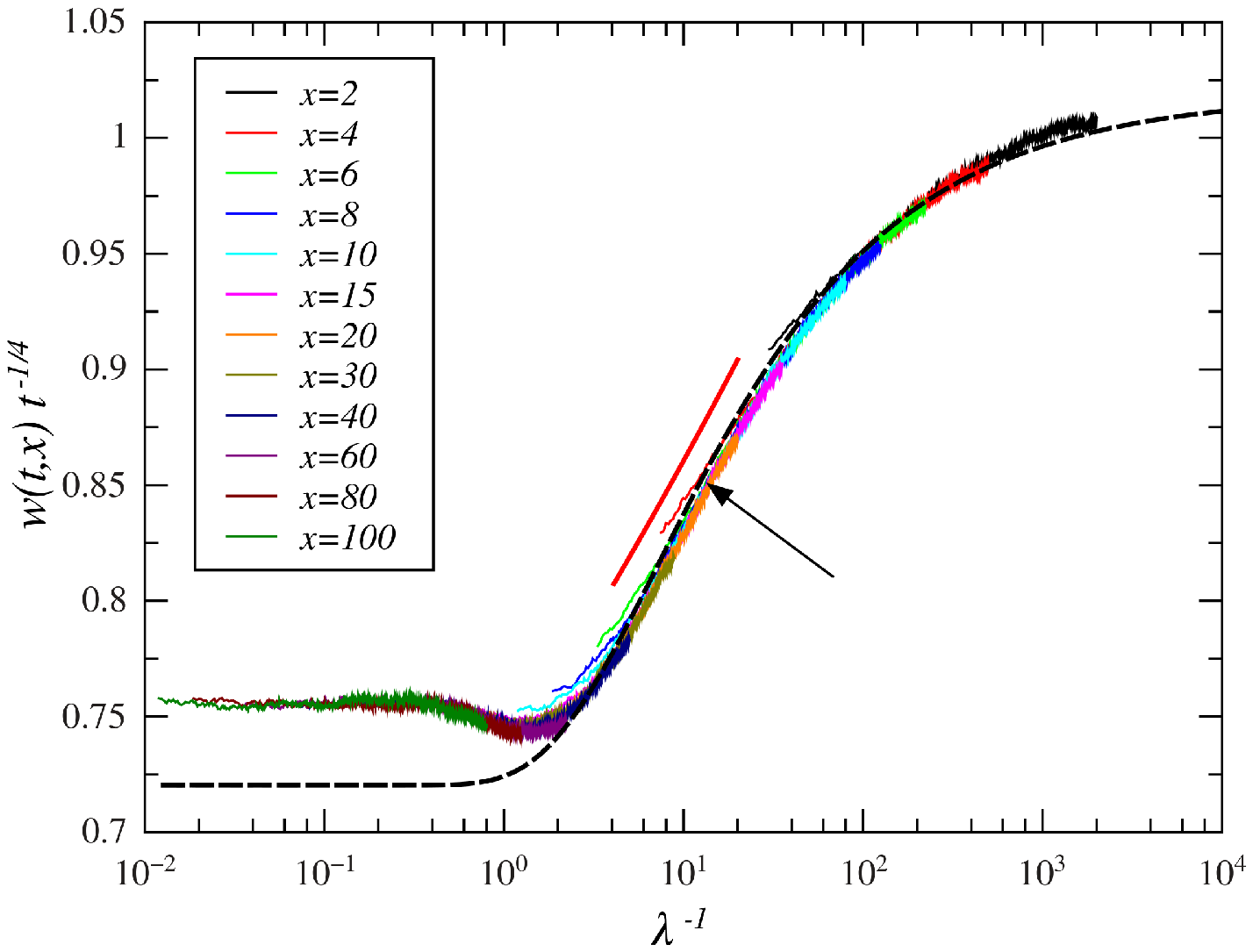}
\label{fig2b}
}
\caption{
(a) Mean height profile $\langle h(t,x)\rangle$ 
in the semi-infinite {\sc{ew}} model, and comparison with {\sc{mc}} 
simulations. 
The full
curve gives the scaling function in the first line of eq.~(\ref{eq7}).
(b) Comparison between {\sc{mc}} simulations and 
theoretical prediction for the scaling function $\Phi_w(\lambda)$ (dashed black 
lines, rescaled by a global factor), in the semi-infinite {\sc{ew}} model. The 
scaling variable $\lambda^{-1}=4t x^{-2}$. 
The arrow indicates the location of the turning point and the slope of the 
straight line
has the value $0.07$.
}
\end{center}
\end{figure}

This scaling form implies the absence of a stationary height 
profile, in contrast to earlier results \cite{Wolf90},
although deep in the bulk the profile remains stationary. This is a consequence of the
boundary condition $H_0(t)\ge H_1(t)$. 
Note that very close to the boundary, the average height $h_0(t)\sim \sqrt{t}$ 
grows monotonically with $t$ such that the interface grows
much faster near the boundary than in the bulk.\footnote{\textc{The 
noise-averaged eq.~(\ref{eq3}) 
with the externally given boundary condition
$h_0(t)=2\kappa_1\sqrt{t/\pi}$ is a classic in the theory of the diffusion 
equation \cite{Crank} and reproduces (\ref{eq7}).
Similarly, if we had set $\kappa_1=\kappa_2=0$, 
a flat interface $\langle h(t,x)\rangle =0$ would result, in
disagreement with the {\sc{mc}}.}} 
If we had made a scaling ansatz in the noise-averaged version 
of eq.~(\ref{eq3}), 
only the boundary terms as specified in eq.~(\ref{eq3}) can reproduce $h_0(t)$ 
correctly.  
On the other hand, for $x\to\infty$, the profile decays as
$\left\langle h(t,x)\right\rangle \sim x^{-2} \E^{-\lambda}$ towards its value 
deep in the bulk. 
In  Fig~\ref{fig1}, the mean profile eq.~(\ref{eq7}) is shown to agree 
perfectly with direct {\sc{mc}} simulations.
This confirms the correctness of the Langevin equation eq.~(\ref{eq3}). 
A good fit between {\sc{mc}} simulations and 
theoretical predictions is achieved with $\nu=0.79$ 
\textc{and $\kappa_1=0.075$}, 
both for the mean profile and width.

\begin{figure}
\centerline{\epsfxsize=3.0in\ \epsfbox{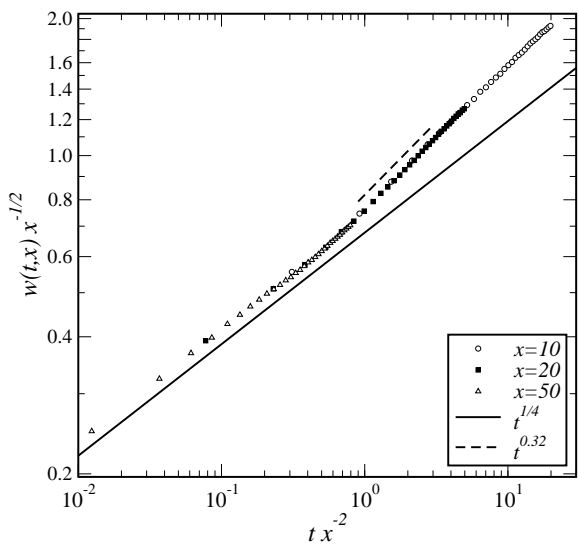}}
\caption[fig2]{Time evolution of the local interface width $w(t,x)$, for several 
distances $x$ from the boundary, 
in the semi-infinite {\sc{ew}} model. A crossover between two regimes is 
clearly visible when time is comparable to the diffusion time.\label{fig2a}}
\end{figure}

Next, we turn to the space-dependent roughness of that interface. In 
Fig~\ref{fig2a}, the time-dependence of the interface width,
as defined in eq.~(\ref{eq1}), is displayed for several distances $x$ from 
the boundary. At small times, one has $w\sim t^{1/4}$ 
(as expected deep inside the bulk \cite{Edwa82,Bara95}) 
which at larger times crosses over towards
$w\sim t^{\beta_{1,{\rm eff}}}$ with an effective exponent which has a greater, non-trivial value 
$\beta_{1,{\rm eff}}\approx 0.32$. 
The scaling $\sim x^2$ of the cross-over time-scale suggests that this cross-over occurs,
at a fixed distance $x$, when causal interactions with the boundary via diffusive transport occur. 
This is analogous to the experimentally observed cross-over to anomalous scaling 
\textc{\cite[Fig.1]{Nasci11}}. 
\textc{Can one take this as evidence in favor of a new {\it `surface 
growth exponent'} 
$\beta_1\ne \beta$, distinct from the
bulk growth exponent $\beta$~? In what follows, we shall show, in spite of the suggestive data, 
that the interpretation of data as in figure~\ref{fig2a} is more subtle.} 

In order to understand this observation, 
we now compute the width analytically, using  
the second line of eq.~(\ref{eq7}). One has
$w^2(t,x) = W_1 + W_2 + W_3$, where, after some computation,
\BEA
W_1 &=& \frac{1}{\pi}\iint_{\lambda}^{\infty}\!\!\!\!\frac{\D v\, \D v'}{\sqrt{v 
v'}\, \E^{v+v'}} \left\langle
\zeta\left( t-\frac{\lambda t}{v},0\right)\zeta\left( t-\frac{\lambda t}{v'},0\right)\right\rangle,
\nonumber \\
W_2 &=& \frac{2}{\sqrt{\pi}} \int_{\lambda}^{\infty} \frac{\D v}{\sqrt{v}\, \E^v} 
\left\langle
\zeta\left( t-\frac{\lambda t}{v},0\right)\zeta\left( t,x\right)\right\rangle,
\\
W_3 &=& \left\langle \zeta^2\left( t,x\right)\right\rangle = 
\frac{T\sqrt{t}}{\sqrt{8\pi}}\int_0^t\!\frac{\D\tau}{\sqrt{\tau}}
\left (1+\erf\sqrt{\frac{2\lambda}{\tau}}\right ).
\nonumber
\EEA
It is easy to see that all three terms can be cast into a scaling form 
$W_{i}=T\sqrt{t}\, {\cal W}_{i}(\lambda)$. ${\cal W}_3$
can be evaluated exactly: $\sqrt{2\pi}{\cal W}_3=
1+\erf\sqrt{2\lambda}+\pi^{-1/2}(2\lambda)^{1/2}\Gamma(0,2\lambda)$, where 
$\Gamma(s,x)$ is an incomplete gamma function. The other integrals can be partially expressed using 
special functions, and the behavior near the boundary can be extracted from 
the series expansion $\lambda\to 0$. We find
\BEA\nn
{\cal W}_1&=&
\frac{1}{\sqrt{2\pi}}+\Big (
\gamma+\ln(2)-2+\ln(\lambda)\Big 
)\frac{\sqrt{\lambda}}{\pi}+O(\lambda^{3/2}),
\\ 
{\cal W}_2&=&
\frac{2}{\sqrt{2\pi}}-2\sqrt{\lambda}+\frac{\pi}{2}\lambda+O(\lambda^2),
\\
{\cal W}_3&=&
\frac{1}{\sqrt{2\pi}}-\Big (\gamma+\ln(2)-2+\ln(\lambda)
\Big )\frac{\sqrt{\lambda}}{\pi}
+O(\lambda^{3/2}).
\nonumber
\EEA
The sum of all these contributions gives $w^2(t,x) =T\sqrt{t}\left 
(4/\sqrt{2\pi}-2\sqrt{\lambda}+O(\lambda^{3/2})\right )$ where the 
logarithmic terms from ${\cal W}_1$ and ${\cal W}_3$ cancel each other. 

In Fig~\ref{fig2b}, the exact scaling function $\Phi_{w}(\lambda) := w(t,x) t^{-1/4}$
is compared with numerical data for a system of size 
$L=10^4$. Clearly, for large $x$ and not too large $t$,
the scaling function is horizontal, 
which reproduces the expected bulk behaviour $w_{\rm bulk}\sim t^{1/4}$. 
However, when the scaling variable $\lambda^{-1}=4tx^{-2}$ 
is increased, the system's behaviour changes such that \textc{
the interface at the boundary is rougher than deep in the bulk, as it is 
exemplified in Fig.~\ref{fig2b} when $\lambda$ is small.} 
For moderate values of $\lambda$, the scaling function
becomes an effective power-law and its slope in Fig~\ref{fig2b} 
can be used to define an effective exponent, here of value $\approx 0.07$.
This reproduces the effective growth exponent 
$\beta_{1,{\rm eff}}=\beta+0.07\simeq 0.32$ observed in Fig~\ref{fig2a}. 
Remarkably, but certainly in qualitative agreement with {\sc{rg}} predictions 
\cite{Lopez05}, the
scaling function does not increase unboundedly as a function of 
$\lambda^{-1}$, but rather undergoes a turning point 
before it saturates, for sites very close to the boundary (where $\lambda\to 0$), 
such that one recovers the scaling $w\sim t^{1/4}$, with a
modified amplitude, however. 
The Langevin equation with boundary terms, eq.~(\ref{eq3}), 
captures completely the change towards the complex
behaviour at intermediate values of $\lambda$ and the saturation in the 
$\lambda\to 0$ limit, but does not yet contain
sufficient detail to follow in complete precision the 
passage from the deep bulk behaviour towards the intermediate regime. 
In any case, the interface growth as described by the semi-infinite {\sc{ew}} 
model is not described by a new
surface exponent but rather by an intermediate regime with an effective 
anomalous growth in a large time window \textc{(between typically 
$1<\lambda^{-1}<10^5$ in Fig.~\ref{fig2b}), before} the standard {\sc{fv}} 
scaling eq.~(\ref{eq2}) with 
$\beta=\frac{1}{4}$ is recovered \textc{but with a higher amplitude}.
\\

\hspace{-0.5truecm}\underline{\it Kardar-Parisi-Zhang model.} 
The simplest non-linear growth model, the paradigmatic $1d$ {\sc{kpz}} equation 
\cite{Kard86}
\BEA \nn
\left(\partial_t -\nu \partial_x^2\right)h(t,x)=\frac{\mu}{2}\left[
\partial_{x}h(t,x)\right]^2+\eta(t,x),
\EEA
is known to describe a wide range of phenomena \cite{Halp95,Krug97,Bara95} 
and admits the exact critical exponents
$z=3/2$, $\beta=1/3$ and $\alpha=1/2$. 
The $1d$ {\sc{kpz}} equation is also known to be exactly solvable 
\cite{Cala11,Sasa10,Sasa10b} and  
recently, an extension to a semi-infinite chain has been studied \cite{Gueu12}. 
Here, we report results of {\sc{mc}}
simulations, based on the {\sc{rsos}} model \cite{Kim89,Henk12} and scaling 
arguments.
The {\sc{rsos}} process uses a integer height variable $H_i(t)\ge 0$ attached 
to 
the sites $i=1,\ldots,L$ of
a linear chain and subject to the constraints
$|H_i(t)-H_{i\pm 1}(t)| \leq 1$, at all sites $i$. It is well-known
that this process belongs to the {\sc{kpz}} universality class and a continuum 
derivation of the 
{\sc{kpz}} equation can be also done as in the {\sc{ew}} case \cite{Park95}.


\begin{figure}[tb]
\begin{center}
\subfigure[ ]
{
\includegraphics[height=5.5cm,keepaspectratio,clip]{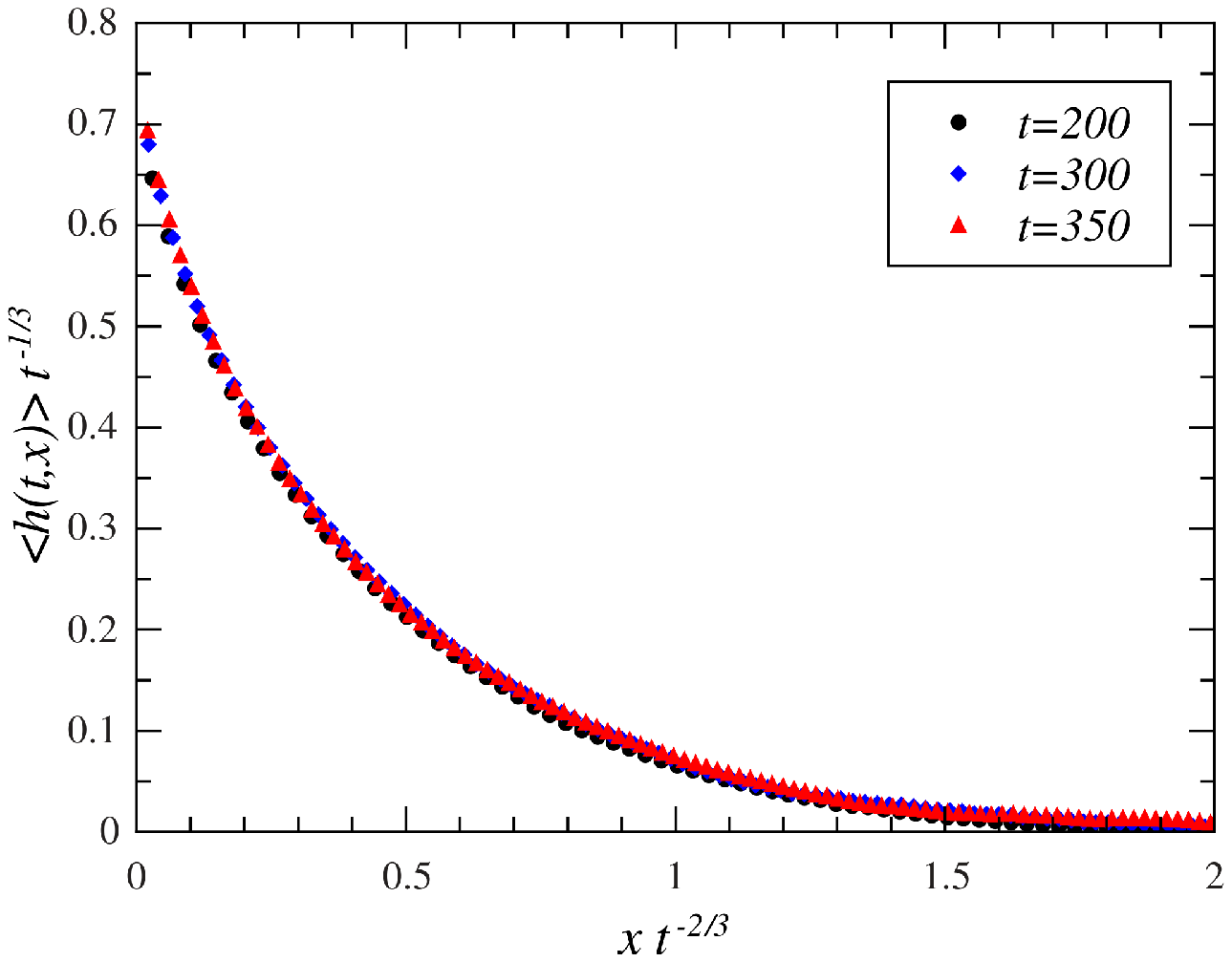}
\label{fig4}
}
\subfigure[ ]
{
\includegraphics[height=5.5cm,keepaspectratio,clip]{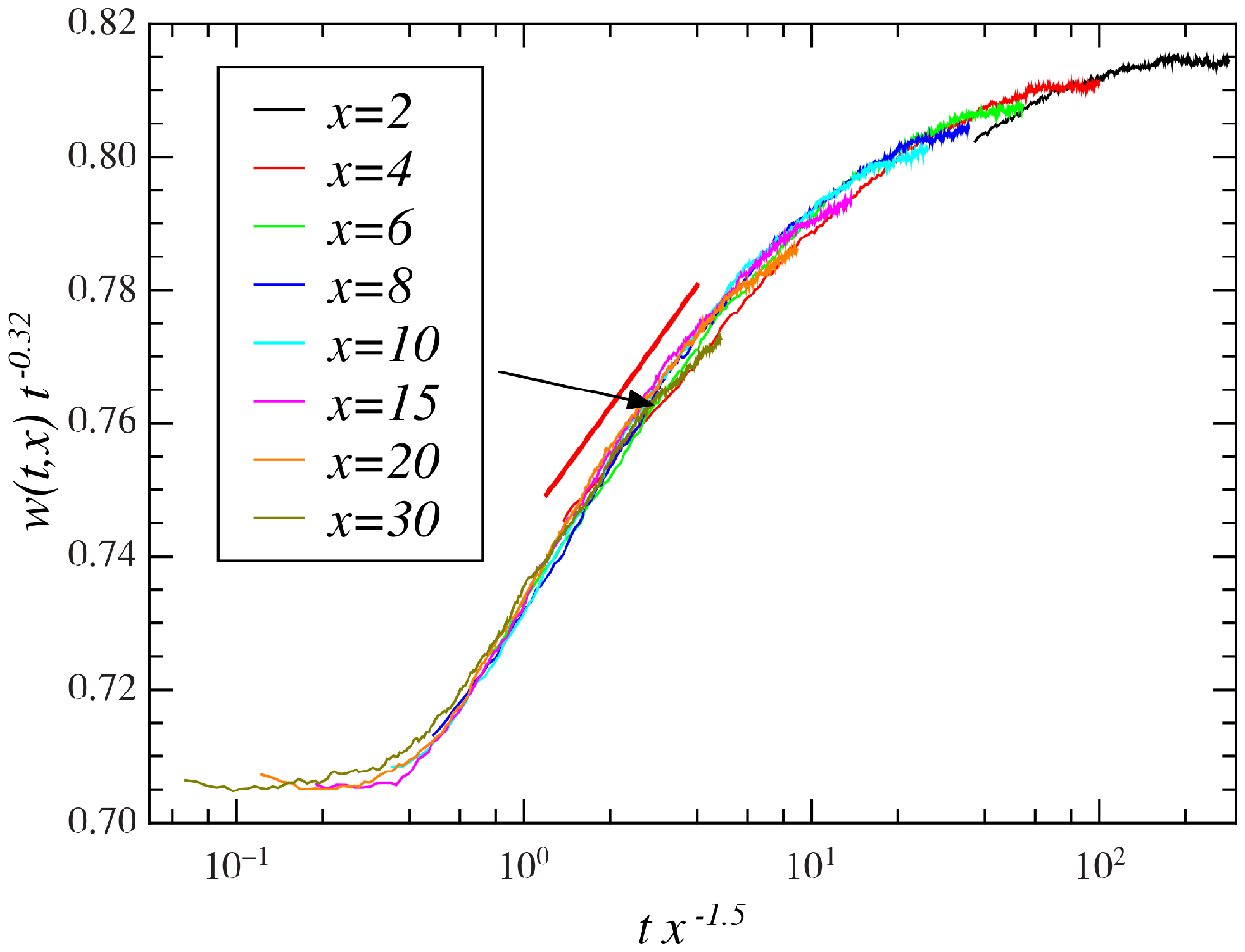}
\label{fig5}
}
\caption{
(a) Height profile $\langle h(t,x)\rangle$ in the semi-infinite {\sc{rsos}} 
process, showing the scaling form (\ref{10}), with $z=\frac{3}{2}$ and 
$\gamma=3$.
(b) Time evolution of the local interface width $w(t,x)$, for several 
distances $x$ from the boundary, 
in the semi-infinite {\sc{rsos}} model.  
The arrow indicates the location of the turning point and the slope of the 
straight line
has the value $0.03$. 
}
\end{center}
\end{figure}

We now introduce a boundary in the {\sc{rsos}} lattice model, 
with modified rules in order to simulate a wall, \textc{see 
fig.~\ref{fig_proc}(ii)}. To this end, we impose $H_0(t)\ge H_1(t)$ such that 
the {\sc{rsos}} condition is always satisfied on the left side of the wall, 
and a particle is deposited on site $i=1$ if the {\sc{rsos}} 
condition $|H_1(t)-H_2(t)| \leq 1$ is fulfilled on the right side.
In our {\sc{mc}} simulations, $L = 10^{3}$ and all the data
have been averaged over $7\cdot 10^{5}$ samples.
The scaling approach is based on the phenomenological boundary-{\sc{kpz}} 
equation in the half-space 
\BEQ\label{eq9}
\left(\partial_t -\nu \partial_x^2\right)h(t,x)-\frac{\mu}{2} (\partial_{x}h(t,x))^2-\eta(t,x)
= \nu\left(\kappa_1 + h_{1}(t)\right)\delta(x). 
\EEQ
In general, one expects a scaling of the profile
\BEQ \label{10}
\langle h(t,x)\rangle =  t^{1/\gamma}\,\Psi(xt^{-1/z}).
\EEQ
In the linear case $\mu=0$, the above exact {\sc{ew}}-solution (\ref{eq7}) 
gives $\gamma=z=2$.
Using the scaling form (\ref{10}) in eq.~(\ref{eq9}), and assuming a mean-field approximation 
$\langle (\partial_{x}h(t,x))^2\rangle\approx (\partial_{x}\langle h(t,x)\rangle)^2$, one may
follow \cite{Bara95} and argue that that the nonlinear part should dominate over the diffusion part.
\textc{Then, we find (the second of these two equations holds true only for 
$\mu\ne 0$)}
\BEA\label{eq10}
\frac{1}{z}+\frac{1}{\gamma}=1 \;\; , \;\;
\frac{2}{z}-\frac{1}{\gamma}=1.
\EEA
For the {\sc{kpz}} class, this implies $z=3/2$ \cite{Kard86} and 
$\gamma=3$ \footnote{This value of $z$ implies a non-diffusive transport
between the bulk and the boundary.}. 
In Fig~\ref{fig4}, the scaling of the profile of the boundary {\sc{rsos}} model 
is shown. The predicted
exponents lead to a clear data collapse, and the shape is qualitatively similar 
to the one of the {\sc{ew}}-class in Fig~\ref{fig2a}.  
The first relation (\ref{eq10}) should be correct for any non-linearity 
describing a boundary growth process because it depends
only of the r.h.s. of eq.~(\ref{eq9}). Hence 
\BEA \label{12}
\gamma=\frac{z}{z-1}=\frac{\alpha}{\alpha-\beta},
\EEA
should be an universal relation for any $1d$ growth process in presence 
of a wall \footnote{For conserved deposition, 
one has the Mullins-Herring equation (see \cite{Bara95}) with $z=4$, 
and eq.~(\ref{12}) gives $\gamma=\frac{4}{3}$, 
which we checked numerically.}.
Turning to the local width, our {\sc{mc}} simulations give again a 
site-dependent behaviour, with a
crossover to an effective exponent $\beta_{1,\rm{eff}}\approx 0.35$, larger than the bulk exponent 
$\beta\approx 0.32$ \footnote{Deviations from the exact theoretical value  $\beta=1/3$ \cite{Kard86}
are due to finite-size effects.}.
The scaling form $w(t,x) t^{-\beta}$ shown in Fig~\ref{fig5} 
displays the same qualitative features as seen before in the
{\sc{ew}} model. This exemplifies that effective anomalous growth behaviour may 
appear in non-linear 
(but non-disordered and local) growth processes.

In summary, in several semi-infinite lattice models of 
interface growth, 
the simple boundary condition $H_0(t)\ge H_1(t)$ on the heights on the two 
sites nearest to the boundary 
not only leads to non-constant and non-stationary height profiles but also to site-dependent roughness
profiles. \textc{There exists a large range of times where} 
effective growth exponents $\beta$ with values clearly larger than in deep in the
bulk can be identified, in qualitative analogy with known experiments on growing interfaces 
\textc{\cite{Nasci11,Yim09,Duerr03,Mata08,Gong11,Kim10,Cord09}}. 
\textc{However, since we have concentrated on models 
defined in the simple geometry of a semi-infinite line with a single boundary, a 
quantitative
comparison with the experiments, 
carried out on faceted growing surfaces with many interacting interfaces, may be 
premature.} 
For non-disordered models with local
interactions, the truly asymptotic growth exponents return to the simple bulk values, 
as predicted by the {\sc{rg}} \cite{Lopez05}. 
The unexpectedly complex behaviour at intermediate times is only seen if appropriate boundary terms 
are included in the Langevin equation describing the growth process. 
Our results were obtained through the exact solution of the semi-infinite 
{\sc{ew}} class and through extensive {\sc{mc}} simulations of both profiles 
and widths in the 
{\sc{ew}} and {\sc{kpz}} models. A 
scaling relation (\ref{12}) for the surface profile exponent $\gamma$ 
was proposed and is in agreement with all presently known
model results. 
\\~\\
{\small This work was partly supported by the Coll\`ege Doctoral 
franco-allemand Nancy-Leipzig-Coventry
(Syst\`emes complexes \`a l'\'equilibre et hors-d'\'equilibre) of UFA-DFH.\\ }


\end{document}